\documentclass[prl, twocolumn, amsmath, nofootinbib, floatfix, showpacs, preprintnumbers]{revtex4}

\usepackage{graphicx}
\usepackage{natbib}
\usepackage{color,comment,aas_macros}

\def\be{\begin{equation}}
\def\ee{\end{equation}}
\def\bea{\begin{eqnarray}}
\def\eea{\end{eqnarray}}
\newcommand{\vs}{\nonumber\\}

\begin{document}

\preprint{LA-UR-11-11572}

\title{Using CMB lensing to constrain the multiplicative bias of cosmic shear}

 \author{Alberto Vallinotto}
 \email{avalli@lanl.gov}
 \affiliation{T-2, MS B285, Los Alamos National Laboratory, Los Alamos, NM 87545}

\date{\today}

\begin{abstract}
Weak gravitational lensing is one of the key probes of cosmology. Cosmic shear surveys aimed at measuring the distribution of matter in the universe are currently being carried out (Pan-STARRS) or planned for the coming decade (DES, LSST, EUCLID, WFIRST). Crucial to the success of these surveys is the control of systematics. In this work a new method to constrain one such family of systematics, known as \textit{multiplicative bias}, is proposed. This method exploits the cross-correlation between weak lensing measurements from galaxy surveys and the ones obtained from high resolution CMB experiments. This cross-correlation is shown to have the power to break the degeneracy between the normalization of the matter power spectrum and the multiplicative bias of cosmic shear and to be able to constrain the latter to a few percent.


\end{abstract}

\maketitle

\textit{Introduction}. Cosmic shear probes the distribution of matter in the universe by measuring the distortions it induces in the ellipticities of background galaxies. In the past decade, this technique has emerged as one of the key probes for cosmology. Thanks to large galaxy surveys that are currently under way (Pan-STARRS \cite{PanStarrs}) or in the works (DES \cite{DES} , LSST \cite{LSST}, EUCLID, WFIRST) weak lensing promises to tightly constrain the large scale properties of the universe and to probe the nature of dark energy to unprecedented accuracy \cite{Hu:1998az, Hu:1999ek, Huterer:2001yu, Refregier:2003ct, Heavens:2003jx, Song:2004tg, Hu:2002rm, Takada:2003ef, Takada:2003hy, Benabed:2003pb, Jain:2003tba, Zhang:2003ii, Bernstein:2003es, Caldwell:2009ix, Frieman:2008sn}. 

Because systematics are one order of magnitude larger than the cosmological signal, their control is critical for this experimental program.
Several different algorithms exist to correct for the instrumental distortions of the point spread function and (in the case of ground based surveys) for the atmospheric seeing \cite{Heymans:2005rv, Massey:2006ha}. As shown by the Shear Testing Programme (STEP) \cite{Heymans:2005rv, Massey:2006ha}, while the application of these algorithms corrects for most systematics, it can introduce bias in the data. Such biases have been catalogued into three generic families: multiplicative and additive biases in shear measurements and errors in redshift measurements \cite{Heymans:2005rv, Massey:2006ha, Huterer:2005ez, Amara:2007as}. The focus of the present work is \textit{multiplicative bias}. The lack of precise knowledge of multiplicative bias leads to a dramatic degradation in the accuracy of the cosmological parameters thus measured \cite{Huterer:2005ez, Amara:2007as}: for example, lack of percent-level constraints on multiplicative bias can lead to an increase in the errors in the value of $\Omega_m$ and $\sigma_8$ of 100\% or more \cite{Huterer:2005ez}. Quite generally, multiplicative bias is insidious because it does not show any scale dependence that can be exploited to decouple it from the weak lensing signal. As such, it is completely degenerate with the normalization of the matter power spectrum: if the observed ellipticities were the only information available it would always be possible to trade a non-zero multiplicative bias for a variation in the total matter in the universe responsible for the lensing signal. A first solution to this problem was proposed in Vallinotto et al.~\cite{Vallinotto:2010qm}, where galaxies' size and luminosity information is used to constrain this bias. Such a solution, however, relies on somehow quantifying the impact of weak lensing on the populations of galaxy sizes and luminosities. 

The present work proposes an alternative way to constrain shear multiplicative bias by cross-correlating it with the \textit{weak lensing} of the Cosmic Microwave Background (CMB). Intuitively, these two probes measure essentially the same observable (albeit extending over different redshift ranges) but using two completely different techniques. Because the systematics affecting these two techniques have completely different natures, cross-correlating these two signals is an effective way to constrain their impact.

\textit{The CMB Lensing Field}.
Experiments aimed at measuring small scale temperature and polarization fluctuations of the CMB \cite{CMB_Experiments} have recently reached the sensitivity required to reconstruct  the effective deflection field arising from the dark matter structures present between the last scattering surface and the observer. CMB lensing has first been detected by WMAP \cite{Smith:2007rg, Hirata:2008cb} and recent measurements from ACT have reported the first detection of its power spectrum \cite{Das:2011ak}. 
The measurement of the lensing of CMB through quadratic optimal estimators \cite{Hu:2001kj, Okamoto:2003zw, Hirata:2003ka} exploits the statistical properties of the primary CMB anisotropies. The dominant systematics for this measurement are the ones characterizing high resolution CMB experiments: emission from unresolved radio and dusty star forming galaxies and thermal and kinetic Sunyaev-Zeldovich effects \cite{Millea:2011pa}. As such, these systematics are completely uncorrelated from the ones characterizing cosmic shear measurements, which are related to the treatment of galaxy images (in particular, atmospheric seeing and the correction of anisotropies in the point spread function) and to the measurement of their ellipticities.

The lensing convergence $\kappa(\hat{n},\chi_f)$ measured for a source at comoving distance $\chi_f$ along a line-of-sight (los) in the direction $\hat{n}$ \cite{Bartelmann:1999yn, Schneider:2005ka, Lewis:2006fu} is defined as
\begin{equation}
\kappa(\hat{n},\chi_f)\equiv C \int_0^{\chi_f}d\chi\,W_L(\chi,\chi_f)\frac{\delta(\hat{n},\chi)}{a(\chi)},\label{Eq:DefKappa}
\end{equation}
where $C=3\Omega_m H_0^2/(2c^2)$ and $W_L(\chi,\chi_f)=\chi(\chi_f-\chi)/\chi_f$ is the lensing window function. 
High resolution CMB experiments allow the reconstruction of the convergence field $\kappa^{\rm obs}_c(\hat{n},\chi_{\rm LSS})\equiv\kappa^{\rm obs}_c(\hat{n})$, which extends all the way to the last scattering surface. 
In general, the convergence measured from the CMB can be modeled as
\begin{equation}
\kappa_c^{\rm obs}(\hat{n})=\kappa_c^{\rm true}(\hat{n})+n_c(\hat{n}),
\end{equation}
where an additive bias $n_c(\hat{n})$ (or more precisely a set of them \cite{Hanson:2010rp}) can be introduced by the reconstruction procedure \cite{Kesden:2003cc}. On the other hand, the optimal quadratic estimators used to reconstruct the deflection field can be properly normalized \cite{Hanson:2010rp}, so as not to introduce multiplicative bias in the CMB lensing signal. In what follows a cutoff on the angular modes contributing to the CMB lensing signal is introduced. This cutoff is determined so that only multipoles that can be measured with signal-to-noise ratio greater than unity are included. In this (imaging) regime the noise term $n_c$ can be neglected. 
\begin{table}[t]
\begin{center}
\begin{tabular}{l|c|c|c|cc}
\hline
Survey & Pixel size & Galaxies per& Sky coverage & $\sigma_{s,N}$ \\
 & (sq. arcmin.) &  $\kappa_s^{\rm obs}$ pixel & (sq. deg.) &\\
\hline
\hline
DES  & 10 & 15  & 5000 &  0.3\\
LSST &10 & 100 & 20000 & 0.3\\
\hline
\end{tabular}
\end{center}
\caption{\label{tab:cosmicshear}
Parameters assumed in the calculation for the two cosmic shear surveys. $\sigma_{s,N}$ is the rms of $\kappa_s$ in the absence of signal, due to shape noise and measurement errors for a single galaxy.} 
\end{table}

\textit{Cosmic Shear and Multiplicative Bias.} Weak lensing surveys measure cosmic shear in redshift bins of finite thickness. The convergence they aim at measuring is
\begin{equation}
\kappa_s^{\rm true}(\hat{n},z_0)=\int_0^{\infty}d\chi \, \eta(\chi) \,\kappa(\hat{n},\chi),
\label{Eq:DefKs}
\end{equation}
where $z_0$ denotes the center of a redshift bin of thickness $\Delta z$ and $\eta(\chi)$ is a selection function for the given redshift bin, normalized so that $\int_0^{\infty}d\chi \eta(\chi)=1$. The convergence value measured from cosmic shear data in a given redshift bin will in general differ from the ``true'' value because of the (possibly redshift dependent) multiplicative bias, so that
\begin{equation}
\kappa_s^{\rm obs}(\hat{n},z_0) = b(z_0) \,\kappa_s^{\rm true}(\hat{n},z_0).
\end{equation}
In what follows a single redshift slice with $z\in[0.9,1]$ is considered and on such redshift slice the multiplicative bias is assumed to be constant. It is straightforward to extend this analysis to other redshift ranges.

\textit{Correlation calculation.} Given a pair of surveys, the data set will consist of the observed $\{\kappa^{\rm obs}_c, \kappa^{\rm obs}_s\}$ over the patch of sky where the two surveys overlap. From Eqs.~(\ref{Eq:DefKappa}-\ref{Eq:DefKs}), it is straightforward to obtain expressions for the elements appearing in the covariance matrix of the joint $\{\kappa^{\rm obs}_c,\kappa^{\rm obs}_s\}$ data set. Let $\hat{n}_i$ and $\theta_{ij}$ denote the direction of the i-th pixel and the angular separation between pixels directed along $\hat{n}_i$ and $\hat{n}_j$ respectively. Furthermore, let $\alpha$ be the matter power spectrum normalization, so that $P(k)=\alpha^2\mathcal{P}(k)$, where $\mathcal{P}(k)$ denotes the shape of the power spectrum. Then, defining
\begin{eqnarray}
 g(\chi)&\equiv&\int_{\chi}^{\infty}d\chi_1 W_L(\chi,\chi_1)\eta(\chi_1),\\
 \zeta(\chi,\theta,\bar{l})&\equiv&\int_0^{\infty}\frac{l\,dl}{2\pi\chi^2}\,J_0(l\theta)
 \,e^{-\left(l/\bar{l}\right)^2}
 \alpha^2\,\mathcal{P}\left(\frac{l}{\chi},\chi\right),
\end{eqnarray}
and using Limber's approximation, the correlations between the different data sets (denoted for brevity by $\langle\kappa^{\rm obs}_{\mu}(\hat{n}_i)\kappa^{\rm obs}_{\nu}(\hat{n}_j)\rangle\equiv\langle\kappa_{\mu}\kappa_{\nu}\rangle_{ij}$ with $\{\mu,\nu\}=\{c,s\}$) are given by the following expressions
\begin{eqnarray}
\langle\kappa_c\kappa_s\rangle_{ij} &=& C^2\,b\int_0^{\infty}d\chi\frac{g(\chi)\,W_L(\chi,\chi_{\rm LSS})}{a^2(\chi)}\zeta(\chi,\theta_{ij},\bar{l}_{cs}),\vs\label{Eq:kcks}\\
\langle\kappa_c\kappa_c\rangle_{ij}&=&C^2\int_0^{\infty}d\chi\frac{W_L^2(\chi,\chi_{\rm LSS})}{a^2(\chi)}\zeta(\chi,\theta_{ij},\bar{l}_{cc}),\\
\langle\kappa_s\kappa_s\rangle_{ij}&=&C^2\,b^2\int_0^{\infty}d\chi\frac{g^2(\chi)}{a^2(\chi)}\zeta(\chi,\theta_{ij},\bar{l}_{ss})+\delta_{ij}\,\sigma_{s,N}^2,\vs\label{Eq:ksks}
\end{eqnarray}
where $\bar{l}_{\mu\nu}\equiv l_{\mu} \, l_{\nu}/(l_{\mu}^2 + l_{\nu}^2)^{1/2}$. Here $\sigma_{s,N}$ is the rms of the shear measurement in absence of signal -- due to the intrinsic ellipticities of galaxies and measurement errors (scaling as $N^{-1/2}$, where $N$ is the average number of galaxies in a $\kappa_s$ pixel) -- and $l_s$ is defined as the limiting multipole corresponding to the $\kappa_s$ pixel size. Similarly, $l_c$ represents the cutoff on the modes contributing to $\kappa_c$. As previously mentioned, $l_c$ is fixed to be the limiting multipole where the signal-to-noise ratio is equal to one. 
\begin{table}[t]
\begin{center}
\begin{tabular}{l|c|cccc}
\hline
Survey & $l_c$ & Sky coverage \\
 &  &  (sq. deg.) \\
\hline
\hline
Planck           & 100 & All sky \\
WidePol & 300 & 4000 \\
CMBPol	& 1000 & All sky\\
\hline
\end{tabular}
\end{center}
\caption{\label{tab:cmblens} Parameters assumed in the calculation for the CMB lensing surveys.} 
\end{table}
\begin{figure*}[t]
\includegraphics[width=0.49\textwidth]{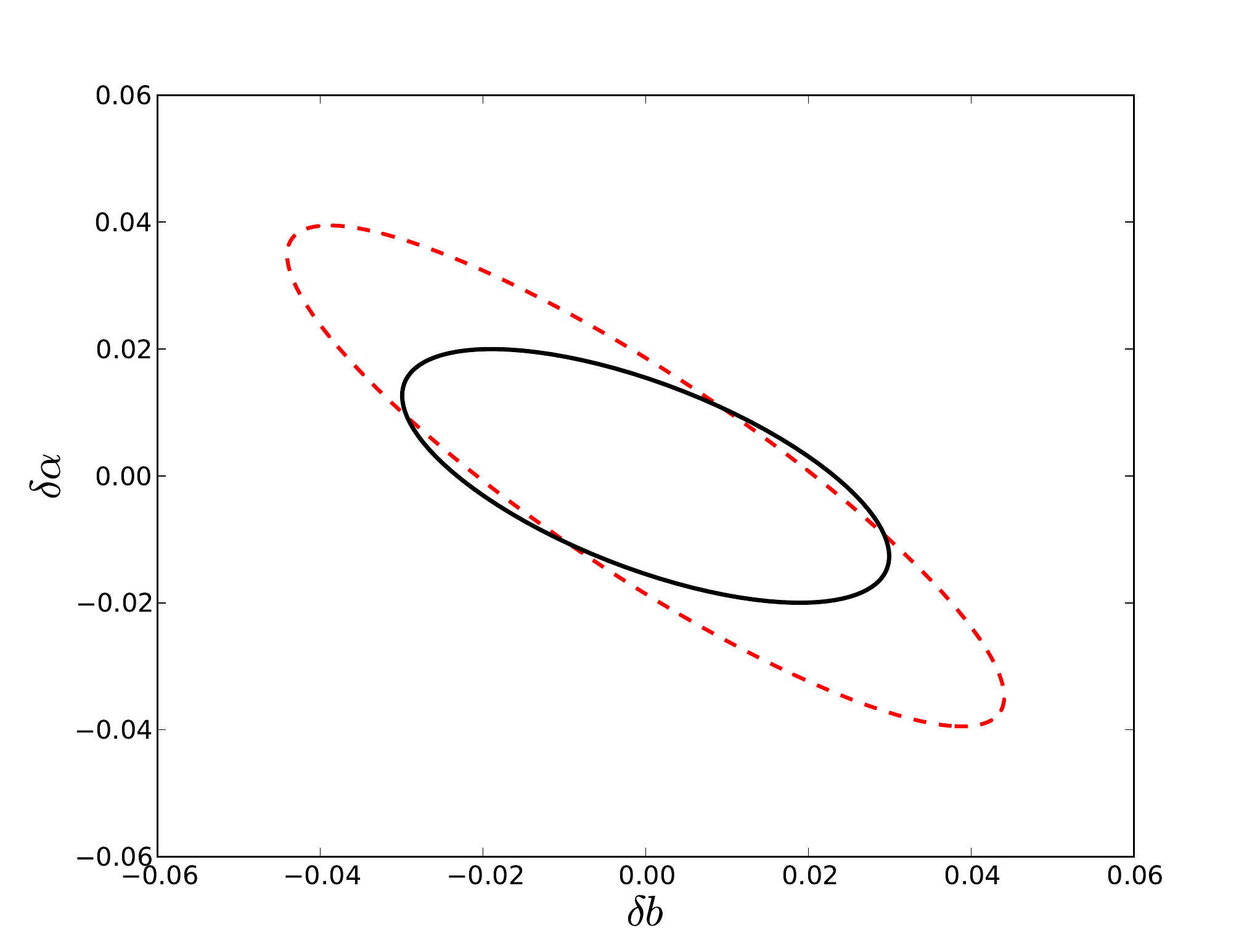}
\includegraphics[width=0.49\textwidth]{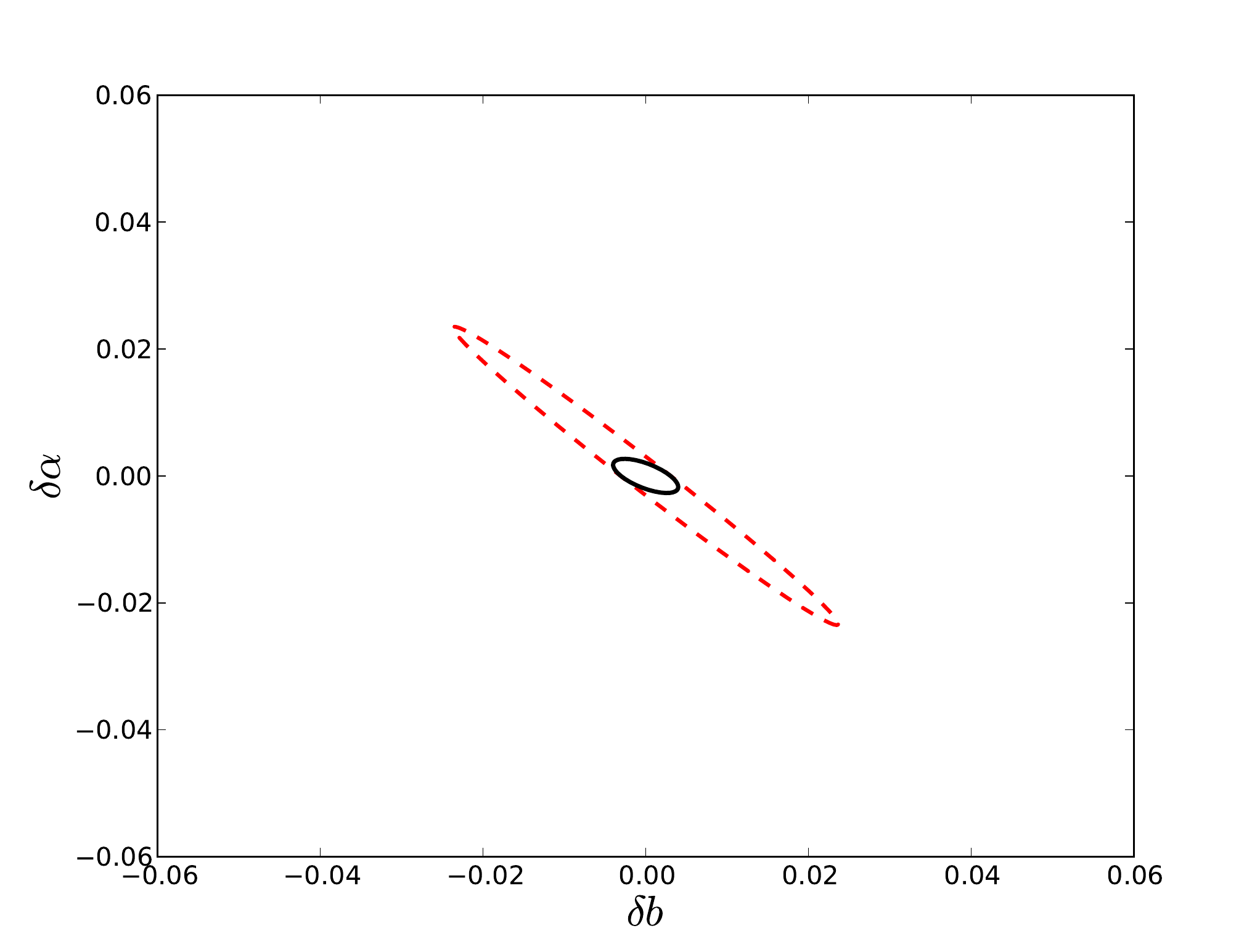}
\caption{\label{fig:DES+LSST} Projected constraints on multiplicative biases and on the normalization of the matter power spectrum $\alpha$ obtained by cross-correlating cosmic shear data from DES (left panel) and LSST (right panel) with CMB lensing data. The red dashed curves show results for cross-correlating with Planck data and include a 4\% prior on the value of $\alpha$. The solid black curve shows results for DES+WidePol (left) or LSST+CMBPol (right) cross-correlations. In the former case, complete overlap between the experiments' footprints is assumed.}
\end{figure*} 

\textit{Constraining Multiplicative Bias.} The Fisher information matrix is used to estimate the constraining power of coupling the $\kappa_s$ and $\kappa_c$ data sets. For the $\kappa_s$ data set, estimates are provided for two surveys: DES and LSST. For the $\kappa_c$ data set, three different surveys are considered: 
an all-sky survey with low resolution reconstruction (Planck \cite{PlanckBB}, $l_c=100$), a futuristic all-sky CMB polarization experiment leading to high resolution reconstruction (CMBPol \cite{2009AIPC.1141....3B}, $l_c=1000$) and a ground based polarization survey -- analogous to ACTPol-Wide \cite{2010SPIE.7741E..51N} and SPTPol \cite{2009AIPC.1185..511M} -- covering 4000 sq.~deg.~at medium resolution (WidePol, $l_c=300$).
The relevant parameters are summarized in Tables \ref{tab:cosmicshear} and \ref{tab:cmblens}.

Given a pair of surveys, the data sets will consist of the $M$ values $\{\kappa^{\rm obs}_{c,i}\}$ and of the $N$ values$\{\kappa^{\rm obs}_{s,j}\}$. In general, the Fisher matrix is
\begin{equation}
F_{\lambda\phi}=\frac{1}{2}{\rm Tr}\left[C_{,\lambda}C^{-1}C_{,\phi}C^{-1}\right],\label{Eq:Fisher}
\end{equation}
where $\lambda$ and $\phi$ run over the parameters $\{\alpha,b\}$, the trace is over all the $M+N$ observables and $C$ is the data covariance matrix with elements given by Eqs.~(\ref{Eq:kcks}-\ref{Eq:ksks}). Because of the very large number of $\kappa_s^{\rm obs}$ pixels ($1.8\times10^6$ for DES and $7.2\times10^6$ for LSST), the evaluation of the Fisher matrix poses a numerical challenge. 
It is however possible to obtain a reasonable (and conservative) estimate of the Fisher matrix by noting that all correlation functions decrease significantly with increasing angular separation. In what follows regions of $(3.6)^2$ sq.~deg.~are treated as statistically independent: over this range of separations, the correlations between the different pixels drop by more than two orders of magnitude.

The left panel of Fig.~\ref{fig:DES+LSST} shows the constraints on $\{\alpha,b\}$ projected for DES+Planck (red dashed contour) and for DES+WidePol (black solid contour). In general, cross-correlating cosmic shear and CMB lensing allows to break the degeneracy between the normalization of the matter power spectrum and cosmic shear's multiplicative bias and to constrain the latter. Because of the limited number of $\kappa_c^{\rm obs}$ pixels that can be reconstructed using Planck data (assuming the conservative value of $l_c=100$), for DES+Planck and LSST+Planck a 4\% prior on $\alpha$ (consistent with current constraints on $\sigma_8$) is assumed to improve the constraints. In the DES+WidePol case, on the other hand, complete overlap between the experiments' footprints is assumed. Under these assumptions the Fisher matrix estimates show that DES+Planck (DES+WidePol) data should allow to constrain shear multiplicative bias $b$ to about 4\% (2\%). The left panel of Fig.~\ref{fig:DES+LSST} can be directly compared to Fig.~2 of \cite{Vallinotto:2010qm}, showing that this cross-correlation represents a viable alternative to using sizes and luminosity information, free from effective parameters modeling the impact of lensing on the size and luminosity populations.

The right panel of Fig.~\ref{fig:DES+LSST} shows projections for LSST+Planck (red dashed contour) and LSST+CMBPol (black solid contour). These data sets benefit from the increase in the cosmic shear footprint and the constraints on $b$ are respectively reduced to 1.7\% and 0.3\%.

\textit{Discussion and conclusions.} The interplay between CMB lensing and cosmic shear data sets allows to lift the degeneracy between the power spectrum normalization and cosmic shear's multiplicative bias and to constrain the latter. This is primarily due to the fact that the CMB lensing kernel is wide and even if it peaks at deeper redshifts, it is broad enough to give a significant non-zero correlation with cosmic shear measurements. 

The results reported thus far are based on the quite remarkable fact that the CMB lensing signal reconstructed from optimal quadratic estimators can be affected by \textit{additive} bias but not by a scale independent multiplicative bias. In other words, as long as optimal quadratic estimators are used, the CMB lensing signal is properly normalized \cite{Hanson:2010rp}. It is however of practical importance to quantify the impact of relaxing such a condition. To do this, the CMB lensing multiplicative bias $c$ is introduced, so that 
\begin{equation}
\kappa_c^{\rm obs}(\hat{n})\equiv c\kappa_c^{true}(\hat{n})+n_c(\hat{n}).
\end{equation}
The resulting enlarged set of multiplicative parameters $\{\alpha,b,c\}$ is degenerate with respect to the cosmic shear and the CMB lensing data sets. It is nonetheless possible to estimate the constraining power of the latter with respect to  $b$ and $c$ by imposing a prior on the normalization of the power spectrum. The results obtained by assuming a 4\% prior on $\alpha$ (consistent with current constraints on $\sigma_8$) are reported in Table \ref{Tab:b_and_c}. Not surprisingly, the results obtained strongly depend on the prior assumed. However, they also show that \textit{under this assumption} the data allow for testing and constraining both multiplicative biases at the few percent level. 

It is furthermore possible to speculate that the cross-correlation between these data sets could also be exploited to constrain the additive biases present in cosmic shear and CMB lensing measurements: while CMB lensing and cosmic shear are characterized by additive biases, their cross-correlation is free from these contaminations. These developments are left for future investigation.

Finally, it is possible to remark that the strong synergy between these different probes of the dark matter distribution suggests that future CMB lensing and cosmic shear surveys should greatly benefit from sharing the same footprint on the sky. This fact is not restricted to cosmic shear surveys but it applies in general to most astrophysical surveys (Lyman-$\alpha$ \cite{Vallinotto:2009wa, Vallinotto:2009jx}, BAO, 21-cm): because CMB lensing is only sensitive to the dark matter distribution, free from any biasing relation, the cross-correlation of its signal with other astrophysical observables allows to extract biasing relations and to control for systematics.

\begin{acknowledgments}
AV is supported by DOE at LANL under contract DE-AC52-06NA25396.
It is a pleasure to thank Salman Habib, Katrin Heitmann, Scott Dodelson and Sudeep Das for a number of discussions throughout the development of this project.

\end{acknowledgments}

\begin{table}[t]
\begin{center}
\begin{tabular}{l|c|cccc}
\hline
Data sets & b & c \\
\hline
\hline
DES + Planck           & 4.2\% (3.6\%) & 5.3\% \\
DES + WidePol 	& 4.3\% (2.0\%) & 4.2\% \\
LSST + Planck		& 4.0\% (1.7\%) & 4.3\%  \\
LSST + CMBPol	& 4.0\% (0.3\%) & 4.0\%  \\
\hline
\end{tabular}
\end{center}
\caption{\label{Tab:b_and_c} Fisher matrix estimates of the constraints on the multiplicative biases for cosmic shear measurements $b$ and for CMB lensing reconstruction $c$ obtained from cross-correlating the two data sets. Because the two multiplicative biases are jointly degenerate with the normalization of the power spectrum $\alpha$, a 4\% prior on the latter is assumed, consistent with current measurements of $\sigma_8$. In parenthesis is the constraint value for an analysis not including $c$.} 
\end{table}
%

\bibliography{VDSV_1028}

\end{document}